\documentclass[%
,twocolumn%
,amssymb, amsmath,nobibnotes, aps, pre]{revtex4-1}
\usepackage{bm}%
\expandafter\ifx\csname package@font\endcsname\relax\else
 \expandafter\expandafter
 \expandafter\usepackage
 \expandafter\expandafter
 \expandafter{\csname package@font\endcsname}%
\fi

\usepackage{graphicx,subfigure}
\usepackage{amssymb,amsmath,wasysym}
\usepackage{color}

\begin{document}

\title{A tug-of-war between stretching and bending in living cell sheets}

\author{P. Recho$^{1,2}$, J. Fouchard$^{3}$, T. Wyatt$^{3,4}$, N. Khalilgharibi$^{3,4}$, G. Charras$^{3,5,6}$, A. Kabla$^{2}$.}

\affiliation{
$^1$LIPhy, CNRS--UMR 5588, Universit\'e Grenoble Alpes, F-38000 Grenoble, France\\
$^2$Department of Engineering, Cambridge University, Cambridge, UK\\
$^3$London Centre for Nanotechnology, University College London, London, UK\\
$^4$Centre for Computation, Mathematics and Physics in the Life Sciences and Experimental Biology, University College London, London, UK\\
$^5$Institute for the Physics of Living Systems, University College London, London, UK\\
$^6$Department of Cell and Developmental Biology, University College London, London, UK
}
\email{\mbox{pierre.recho@univ-grenoble-alpes.fr} }

\date{\today}

\begin{abstract}

The balance between stretching and bending deformations characterizes shape transitions of thin elastic sheets. While stretching dominates the mechanical response in tension, bending dominates in compression after an abrupt buckling transition. Recently, experimental results  in suspended living epithelial monolayers have shown that, due to the asymmetry in surface stresses generated by molecular motors across the thickness $e$ of the epithelium, the free edges of such tissues spontaneously curl out-of-plane, stretching the sheet in-plane as a result. This suggests that a competition between bending and stretching  sets the morphology of the tissue margin. 
In this study, we use the framework of non-euclidean plates to incorporate active pre-strain and spontaneous curvature to the theory of thin elastic shells. We show that, when the spontaneous curvature of the sheet scales like $1/e$, stretching and bending energies have the same scaling in the limit of a vanishingly small thickness and therefore both compete, in a way that is continuously altered by an external tension, to define the three-dimensional shape of the tissue.

\end{abstract}


\maketitle

\section{Introduction}
Active surfaces are ubiquitous in biology, ranging from sub-cellular organelles to the complex multi-layered walls compartmentalising organs. An important feature of such surfaces compared to classical visco-elastic materials is that their mechanical properties depend of the controlled cellular metabolic processes that continuously inject energy and maintain mechanical tension in the system \cite{lecuit2007cell}. 
Such activity is responsible for the appearance of a cleavage cytokinetic furrow driving the division of a single cell \cite{turlier2014furrow, reymann2016cortical}, or multicellular topological transitions during development such as mesoderm invagination in \textit{Drosophila} \cite{martin2010integration, brodland2010video} and inversion of the \textit{Volvox} embryo \cite{hohn2015dynamics} both involving hundreds of cells. The detailed patterns resulting from these mechanical interactions often involve instabilities  where elastic stretching and bending deformations play an important role, such as \textcolor{black}{in the rupture and subsequent curling of single red blood cells \cite{kabaso2010curling,callan2012red}} or the formation of villi in the gut \cite{shyer2013villification} and gyri and sulci in the brain \cite{tallinen2016growth, holland2018symmetry, karzbrun2018human} which shape entire organs.

\begin{figure}[h!]
\centering
\includegraphics[width=0.4\textwidth]{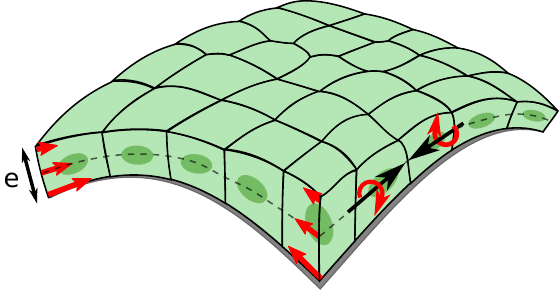}
\caption{\label{fig:scheme_active_film} Sketch of a polar cohesive cell monolayer. Dark ellipses are the cell nuclei. Black lines show cell-cell junctions. The thin grey line indicates the basal side which developed in contact with the substrate while the apical side was free. The substrate is then removed in our mechanical experiments. The apico-basal polarity entails a mechanical polarity stemming from the inhomogeneous distribution of active stress along the thickness (thick red arrows). This results in the presence of active tension along mid-plane (black arrows) as well as active torques out-of plane (rotating arrows).}
\end{figure}

We focus in this paper on epithelial cell monolayers. These tissues are composed of a single layer of cells laterally attached to one another via specialised adhesion proteins, as illustrated on Fig.~\ref{fig:scheme_active_film}. The inner surface of the cells is covered by a thin cortex composed of a dynamic polymer meshwork that can contract thanks to molecular motors which act as active cross-linkers \cite{salbreux2012actin}. Epithelial tissues line the surface of organs and vessels, physiologically defining compartments and regulating transport accros them. As such, epithelial monolayers are polarized, with anatomical differences between the two sides on the monolayer. This includes molecular motors which often exhibit an asymmetric distribution along the sheet thickness axis and are preferentially located to one of the sides \cite{st2011epithelial,asnacios2012mechanics}. In continuum mechanics theories, this uneven distribution of motors gives rise to both active in-plane tensions and out-of-plane torques \cite{liang2009shape, dervaux2009morphogenesis, efrati2009elastic, berthoumieux2014active, murisic2015discrete, krajnc2015theory, salbreux2017mechanics, haas2019nonlinear}.

Our experimental system consists of a suspended cell monolayer devoid of its substrate and clamped between two cantilevers, \textcolor{black}{one fixed and one mobile,} whose spacing can be adjusted (Fig.~\ref{fig:layer}). The protocol is detailed in \cite{harris2012characterizing,harris2013generating}. This experimental condition allows us to specifically probe the mechanical properties of the epithelial sheets in the absence of confounding effects stemming from the substrate.  Since the stiffness of \textcolor{black}{the mobile} cantilever is known, the total traction force on the cantilever can be measured and the active non-linear rheology of such suspended monolayer can be probed across various timescales  \cite{khalilgharibi2019stress, bonfanti2019unified}. We shall focus here on an intermediate timescale ranging roughly between 30 s and 10 min where the macroscopic monolayer stress-strain curve is well captured by an elastic model with an active pre-stress \cite{wyatt2019actomyosin}. However, examining the shape of the free edge of the suspended layer also revealed that the margin locally curls out-of-plane with a high spontaneous curvature of the order of the inverse of tissue thickness \cite{Fouchard806455} (see also Fig.~\ref{fig:layer}~(b) ). 

We therefore adopt the framework of Non-Euclidean elastic Plates (NEP) \cite{efrati2009elastic,pezzulla2017curvature} which generalizes the theory of F\"{o}ppl-von K\'{a}rm\'{a}n \cite{Audoly2010} to account for the presence of both in-plane  pre-strain  as well as spontaneous curvature. The peculiarity of our analysis is that, following experimental observations,  we assume that the spontaneous curvature scales with the inverse of the layer thickness leading to a direct competition between stretching and bending energies to set the shape of the free edge of the monolayer. We then study how this competition is controlled by the active pre-strain and spontaneous curvature using a simple one parameter ansatz which we qualitatively compare to experimental results. Note that we do not aim at quantitatively capturing the experimental results which we rather use to motivate our theoretical study.

The paper is organized as follows.  In section~\ref{sec:kinematics}~ and~\ref{sec:passive_rheo}, we present the elastic film framework that we use to model the passive behavior of the monolayer and exemplify in section~\ref{sec:buckling} \textcolor{black}{that an external compression is needed to observe a buckling behavior characterized by a transition from a stretching to a bending dominated regime}. We then complete our model in section~\ref{sec:active_rheology} by accounting for tissue activity. Next, in section~\ref{sec:one_parameter_reduction}, we assume a simple deformation ansatz that reduces the mechanical problem to a single parameter characterizing the shape of the free edge. \textcolor{black}{We use this parameter  in section~\ref{sec:initial_deflection} to show} that unlike in the passive case, there is a continuous competition between stretching and bending energies even in the absence of external loading. We study in section \ref{sec:deflection_elongation} how applying an external tension changes the balance between stretching and bending, \textcolor{black}{therefore modifying the film free edge.} We finally discuss our results in section~\ref{sec:discussion}.

 \section{Large deformations kinematics of the monolayer}\label{sec:kinematics}
 
We denote $w$ the \textcolor{black}{width of the monolayer at its contact with the cantilever} and  $L$ the separation between the two cantilevers. A displacement $L-L_0$ can be uni-axially imposed from the initial separation $L_0$. 
\begin{figure}[h!]
\centering
\includegraphics[width=0.5\textwidth]{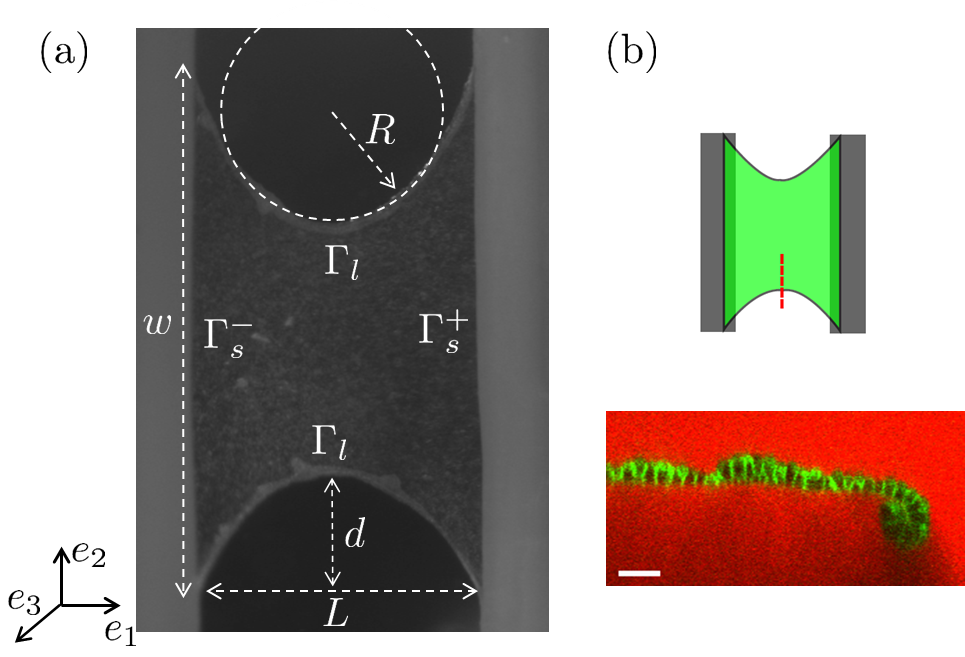}
\caption{\label{fig:layer}  (a) Picture of the cellular layer suspended between two cantilevers. The dimensions of the layer are $L$ and $w=1.6$ mm. The typical (in-plane) curvature of the free edges is denoted by $R$ and the deflection in the center by $d$. The contact lines with the cantilevers are $\Gamma_s^{\pm}$ and the free edges are $\Gamma_l$. (b) Profile of the cellular layer in the ($e_2$,$e_3$) plane where $d$ is measured. Cell-cell junctions are marked in green via ECadherin-GFP. Medium is marked in red. Note the radius of curvature of the curl is \textcolor{black}{on the same order of magnitude} as the layer thickness \cite{Fouchard806455}. Profile is taken at the position indicated by the dashed red line on the diagram above. Scale bar: 30$\mu m$.}
\end{figure}
The bulk of the layer at a given time $t$ is denoted $\Omega$, the free lateral surface (i.e. the tissue margin) $\Gamma_l$, the fixed left side surface $\Gamma_s^-$ and the right side $\Gamma_s^+$.  See Fig.~\ref{fig:layer}~(a).  The free margin of the layer bridges the gap between the two cantilevers. The deflection at the middle of the bridge is denoted $d$.  See Fig.~\ref{fig:layer}~(a).
   
 The  applied displacement may become large (of the order of the size of the initial layer) so we do not impose the restriction of small displacements in the  theory presented below. 
In the lab frame $(e_1,e_2,e_3)$, the displacement  of the monolayer is measured from a flat rectangular reference configuration $\Omega_0=(0,L_0) \times (0,w)$ \textcolor{black}{where all the reference boundaries of the domain $\Gamma_{l,0}$, $\Gamma_{s,0}^{\pm}$ are straight lines. This physically corresponds to the configuration of the cell monolayer in the absence of any internal active stress or external loading.} The position of a material point in the current configuration can thus be written
$$f(x)=x+u(x)+v(x)e_3,$$
where \textcolor{black}{the reference configuration coordinate} $x\in \Omega_0: x_1\in (0,L_0),x_2\in (0,w)$ and $u=u_1 e_1+u_2 e_2$ is the in-plane $(e_1,e_2)$ displacement while $v$ is the out-of-plane displacement. The deformation $F$ is the gradient of $f$:
\begin{equation}\label{eq:deformation}
F=\left(\begin{array}{cc}1+\partial_{x_1}u_1&\partial_{x_2}u_1\\
\partial_{x_1}u_2&1+\partial_{x_2}u_2\\
\partial_{x_1}v&\partial_{x_2}v
\end{array}\right).
\end{equation}
\textcolor{black}{$F$ is not a square matrix  because we consider a two dimensional object embedded in a three dimensional space.} From the deformation, we compute the membrane Cauchy-Green  tensor $C=F^{T}F$ and 
\begin{equation}\label{eq:Cauchy_strain}
E=\frac{C-I}{2}
\end{equation}
is an associated strain tensor ($I$ denotes the identity).

The unit normal to the monolayer reads 
$$n=\frac{Fe_1\wedge Fe_2}{\sqrt{\text{det} C}},$$
where $\wedge$ denote the vector product.  
The local curvature tensor at each point of the monolayer is then defined by $K=n.\nabla^2f$, that is in index notation \cite{ciarlet2005introduction}
\begin{equation}\label{eq:Curvature}
K_{i,j}=\sum_{k=1..3}n_k\partial_{x_ix_j} f_k.
\end{equation}

Having defined the strain and curvature of the film, we now use these two variables to specify its mechanical behavior.

\section{Passive rheology and boundary conditions}\label{sec:passive_rheo}
We describe the \emph{passive} response of the cell monolayer as purely elastic.  Due to the thin film approximation, the stored elastic energy $U$ can  be decomposed into membranal and bending terms \cite{li2016stability}:
$$U[u,v]=\int_{\Omega_0}\left[ u_{\text{s}}(E)+u_{\text{b}}(K)\right] dx_1dx_2.$$
For simplicity, we consider only physically linear elasticity (meaning that the energy functionals neglect the terms that are higher than quadratic in the strain and curvature), isotropic and 3D-incompressible (because the volume of each cell remains constant in the regime tested \cite{harris2012characterizing, wyatt2019actomyosin}). The stretching elastic energy $U$ therefore reduces to the classical Saint-Venant Kirchhoff expression
$$u_{\text{s}}(E)=\frac{Ye}{3}\left[(\text{tr} E)^2+\text{tr}(E^TE)\right],$$
where $Y$ is the 3D Young's elastic modulus and $e$ the thickness of the layer \textcolor{black}{in the reference configuration} (small compared to $L_0$ and $h$). Similarly, the bending energy takes the form,
$$u_{\text{b}}(K)=\frac{Ye^3}{36}\left[(\text{tr} K)^2+\text{tr}(K^TK)\right].$$
Note that retaining only quadratic contributions in the energies (i.e. assuming a linear material) is compatible with considering large deformations (i.e. geometrical non-linearities).

Using the internal energy, we can define the Piola-Kirchhoff stress $N=\partial_E u_{\text{s}}$ and the torque $M=\partial_Ku_{\text{b}}$.  \textcolor{black}{On the reference boundary $\Gamma_{l,0}$, the traction stress as well as the torque vanish}. Note the absence of a work term in the above expression of the potential energy $U$ because there is no surface where a non-zero traction stress is imposed. Locally, it is always the displacement which is imposed \textcolor{black}{on $\Gamma_{s,0}^{\pm}$, i.e. $(L-L_0)e_1$ on $\Gamma_{s,0}^{+}$ and clamped conditions on $\Gamma_{s,0}^{-}$. The local traction stress on $\Gamma_{s,0}^{+}$, $eT=FNF^{T}.e_1$, which is opposite to the one on $\Gamma_{s,0}^{-}$  cannot be imposed with this device.  Instead, we impose a certain displacement such that a target global traction force $T_m=|\Gamma_{s,0}^+|^{-1}\int_{\Gamma_{s,0}^+}T.e_1$ is applied  \cite{wyatt2019actomyosin}}. 

\section{The stretching to bending transition in buckling}\label{sec:buckling}

Before moving to our main results, we re-derive in this section some classical results about \textcolor{black}{the stretching and bending behavior} of a plate in plane strain (i.e. equivalent to a one dimensional beam in the $e_1$ direction) for our specific theoretical setting. 

To do so, we consider the case where $u_2=0$ and $u_1(x_1)$ and $v(x_1)$ do not depend on $x_2$. Then, setting $1+\partial_{x_1}u_1=\Lambda \cos(\phi)$ and $\partial_{x_1}v=\Lambda \sin(\phi)$ the strain and the curvature become scalar quantities:
\begin{equation}
E=\frac{\Lambda^2-1}{2}\text{ and } K=-\Lambda\partial_{x_1}\phi.
\end{equation}
The variable $\Lambda$ therefore represents the stretch along the $x_1$ direction and  $\phi$ represents the angle of the plate with its tangent. Using $\Lambda$ and $\phi$ we can re-express the bending and stretching energies:
$$u_s=\frac{2Ye}{3}\left( \frac{\Lambda^2-1}{2}\right)^2\text{ and } u_b=\frac{Ye^3}{18}\left(\Lambda\partial_{x_1}\phi\right)^2$$ 
such that the total elastic energy reads,
$$U[\Lambda,\phi]=\int_0^{L_0}\left(\frac{2Ye}{3}\left( \frac{\Lambda^2-1}{2}\right)^2+ \frac{Ye^3}{18}\left(\Lambda\partial_{x_1}\phi\right)^2\right) dx_1.$$

In our problem, the plate is clamped  at $x_1=0$ and $x_1=L_0$: $v(0)=v(L_0)=0$, $u(0)=0$ and $u(L_0)=L-L_0$ with a slope that we assume null ($\partial_{x_1}v(0)=\partial_{x_1}v(L_0)=0$).  In the new variables $\Lambda$ and $\phi$, these boundary conditions become the integrals constraints
$$\int_0^{L_0}\left(\Lambda\cos(\phi)-1-\epsilon^m\right)dx_1=0 \text{ and } \int_0^{L_0}\Lambda\sin(\phi)dx_1=0,$$
 where $\epsilon^m=(L-L_0)/L_0$ and the boundary conditions
$$\phi(0)=\phi(L_0)=0.$$

The solution of this problem is therefore obtained by minimization of the Lagrangian
\begin{multline}
\mathcal{L}[\Lambda,\phi]=U[\Lambda,\phi]-P\int_0^{L_0}\left(\Lambda\cos(\phi)-1-\epsilon^m\right)dx_1\\-Q\int_0^{L_0}\Lambda\sin(\phi)dx_1,
\end{multline}
where the Lagrange multipliers $P$ and $Q$ represent the forces at the boundary in the $e_1$ and $e_3$ directions. The first variation of $\mathcal{L}$ provides the two coupled equations  determining  the equilibrium shape:
\begin{equation}\label{e:buckling_pb}
\begin{array}{c}
\frac{Y e^3}{9}\Lambda(\partial_{x_1}\phi)^2+\frac{2 Y e}{3}\Lambda(\Lambda^2-1)=P\cos(\phi)+Q\sin(\phi)\\
\frac{Y e^3}{9}\partial_{x_1}\left(\Lambda^2 \partial_{x_1}\phi \right) =Q\Lambda\cos(\phi)-P \Lambda\sin(\phi).
\end{array}
\end{equation}
When $\epsilon^m>0$ (i.e. the film is put under tension \textcolor{black}{ as in the experimental conditions that we will study in the rest of the paper where the monolayer is enriched with an active behavior}), the solution of \eqref{e:buckling_pb} clearly corresponds to a pure stretching case which is given by (with obvious notations)
$$\phi_s=0\text{, } \Lambda_s=L_0(1+\epsilon^m) \text{, } Q_s=0  \text{ and } P_s=\frac{2Ye}{3}\Lambda_s(\Lambda_s^2-1).$$ 
The total bending energy 
$$U_b=\int_0^{L_0}u_b(x_1)dx_1$$
therefore vanishes and the total stretching energy 
$$U_s=\int_0^{L_0}u_s(x_1)dx_1,$$
scales with $e$ leading to a stretching energy dominated regime where the bending term is irrelevant.

The case $\epsilon^m<0$ \textcolor{black}{(i.e. the film is put under compression)} is more complex and corresponds to a classical buckling problem. When $\epsilon^m$ is lower than the deformation corresponding to the critical loading threshold 
$$\epsilon^m_c=\sqrt{1-\frac{2 \pi ^2 e^2}{3 L_0^2}}-1,$$
the trivial solution $\phi_s, \Lambda_s, Q_s, P_s$ stops to be stable and  bifurcates through a second order phase transition to a non-homogeneous solution which can be expanded in power series close to the bifurcation point using the Lyapunov-Schmidt reduction technique \cite{Amazigo1970, Koiter1976}. Following this approach, the normal form up to second order reads:
\begin{equation}\label{e:power_series_bif}
\begin{array}{c}
\phi_c(x_1)\simeq\phi_s+\nu\phi_1(x_1)+\nu^2\phi_2(x_1)\\
\Lambda_c\simeq\Lambda_s+\nu \Lambda_1(x_1)+\nu^2\Lambda_2(x_1)\\
Q_c\simeq Q_s+\nu Q_1+\nu^2 Q_2\\
P_c\simeq P_s+\nu P_1+\nu^2 P_2.
\end{array}
\end{equation}
As it is classical for a second order phase transition (i.e. a super-critical pitchfork bifurcation), the small parameter in the expansion is given by 
$$\nu=\sqrt{\frac{\epsilon^m-\epsilon_c^m}{\epsilon^m_2}}$$
and in our specific problem,
$$\phi_1(x_1)=\phi_2(x_1)=\sqrt{2} \sin \left(\frac{2 \pi  x_1}{L_0}\right),$$
$$\Lambda_1=P_1=Q_1=Q_2=0$$ 
while 
\begin{multline}
\Lambda_2(x_1)=\\
\frac{\pi ^2 e^2 \left(\epsilon _c^m+1\right) \left(2 \left(\frac{2 \pi ^2 e^2}{L_0^2}-3\right) \cos \left(\frac{4 \pi  x_1}{L_0}\right)+\frac{6 \pi ^2 e^2}{L_0^2}-3\right)}{4 L_0^2 \left(\frac{2 \pi ^4 e^4}{L_0^4}-\frac{5 \pi ^2
   e^2}{L_0^2}+3\right)},
\end{multline}
$$P_2=\frac{\pi ^2 e^3 Y \left(3-\frac{14 \pi ^2 e^2}{L_0^2}\right) \left(\epsilon _c^m+1\right)}{9 L_0^2 \left(\frac{2 \pi ^2 e^2}{L_0^2}-1\right)}$$
and 
$$\epsilon^m_2=\frac{\left(\frac{2 \pi ^4 e^4}{L_0^4}+\frac{7 \pi ^2 e^2}{L_0^2}-6\right) \left(\epsilon _c^m+1\right)}{4 \left(\frac{2 \pi ^4 e^4}{L_0^4}-\frac{5 \pi ^2 e^2}{L_0^2}+3\right)}.$$
Using the above expressions we obtain the scaling of the stretching and bending energies for the buckled solution closed to the critical threshold:
$$U_s\simeq \frac{\pi ^4 e^2 Y \left(27 \delta _{\epsilon }^2+12 \delta _{\epsilon }+4\right)}{54}\frac{e^3}{L_0^3} $$  
and 
$$U_b\simeq \frac{4 \pi ^2 e^2 Y \delta _{\epsilon } \left(2 \delta _{\epsilon }+2 \sqrt{2} \sqrt{\delta _{\epsilon }}+1\right)}{9}\frac{e}{L_0}$$ 
where $\delta_{\epsilon}=\epsilon_c^m-\epsilon_c>0$ and we retained only the dominating term in the expansion in $e/L_0$ of the power series.

This shows that as soon as $\delta_{\epsilon}>0$ (i.e. after the bifurcation from the constant solution), for a slender structure ($e/L_0\ll 1$), the bending energy dominates over the stretching energy.

Classical buckling of a passive elastic slender structure therefore involves a transition from a stretching dominated regime to a bending dominated regime at the bifurcation point. In the stretching dominated regime, the bending energy vanishes; while in the bending dominated regime, it is the stretching energy that is negligible. In the more complex 2D theory where plane strain is not assumed, it is possible that stretching of the film in the $e_1$ direction, as imposed with our experimental device, initiates the formation of wrinkles in the $e_2$ direction \cite{li2016stability}. Indeed, volume conservation implies a certain level of compression which activates the bending energy in that direction. Similar to the classical buckling case presented above, these wrinkles happen through a bifurcation indicating a transition -driven by the external loading- from a regime dominated by the stretching energy to a regime where minimization of the bending energy becomes more favorable. 

In the following sections, we will show that, due to active effects, this situation changes as the film exhibits a tug-of-war between the stretching and the bending energies  \textcolor{black}{ even when the film is put under an external tension}. This is because both energies scale in the same way with respect to the small parameter $e/L_0$ to determine the film shape. 

\section{Incorporation of  the active rheology}\label{sec:active_rheology}

The biological activity in the monolayer here refers to a contractile acto-myosin polymer network generating mechanical tension in the plane of the monolayer. This active tension, combined with the elastic modulus of the monolayer, leads to the emergence of an effective pre-strain that can be controlled by modulating the acto-myosin dynamics  \cite{wyatt2019actomyosin}. In addition to the in-plane component of the tension, an asymmetry of myosin activity across the thickness of the monolayer leads to an active torque, which in turn manifests itself as a spontaneous curvature of the monolayer \cite{st2011epithelial,asnacios2012mechanics}.  

Building on the idea of a stretching and bending decomposition, we therefore speculate that the total  potential energy reads,
\begin{equation}\label{e:elastic_energy}
U[u,v]=\int_{\Omega_0} \left[u_{\text{s}}(E-E_a)+u_{\text{b}}(K-K_a)\right]dx_1dx_2,
\end{equation}
where we suppose that the minimum of the internal energy (i.e. the ground state) is shifted by active effects \cite{klein2007shaping, efrati2009elastic, dervaux2009morphogenesis, pezzulla2017curvature}. In particular, we do not consider here the fact that activity may modify the functional form of the energies $u_{\text{s}}$ and $u_{\text{b}}$ themselves.  This expression of the elastic energy has been justified under the classical Kirchhoff-Love assumptions in the limit of small thickness of a bulk elastic material with embedded pre-strain \cite{efrati2009elastic}. However for a spontaneous curvature of the order of $1/e$  created between the apical and basal side of the cell monolayer, one of the Kirchhoff-Love assumptions (the plane-stress assumption) is no longer verified and we therefore use this form of the elastic energy as an effective way to capture the competition between stretching and bending that we experimentally observed, rather than the one originating from a generic thin film limit. 

We also assume that the active contributions are isotropic in the monolayer plane:
$$E_a=\left(\begin{array}{cc}\epsilon_a&0\\0&\epsilon_a\end{array}\right)\text{ and }K_a=\left(\begin{array}{cc}R_a^{-1}&0\\0&R_a^{-1}\end{array}\right),$$
where $\epsilon_a<0$ is an in-plane contractile pre-strain while $R_a$ is a spontaneous radius of curvature. 

The total potential energy $U$ then needs to be minimized in the proper kinematically admissible field of displacement $(u,v)$  (displacements satisfying the imposed displacements boundary conditions) to solve the problem. To gain some analytical insight, we follow below a more simple single parameter analysis that captures the mononolayer shape.

\section{Parametric model of the tissue margin curling}\label{sec:one_parameter_reduction}

In experiments, we noticed the presence of a strong curling at the tissue margin with more pronounced curling in the center of the margin (of the order of $1/e$) \cite{Fouchard806455}. Our hypothesis is that such curling localized at the tissue margin creates the deflection $d$ by relaxing some bending energy. The deflection however remains finite since this operation costs stretching energy as it leads to stretching in the monolayer tangential to the tissue margin. The deflection is thus a compromise between stretching and bending of the cell monolayer.  

\begin{figure*}
\centering
\includegraphics[width=0.9\textwidth]{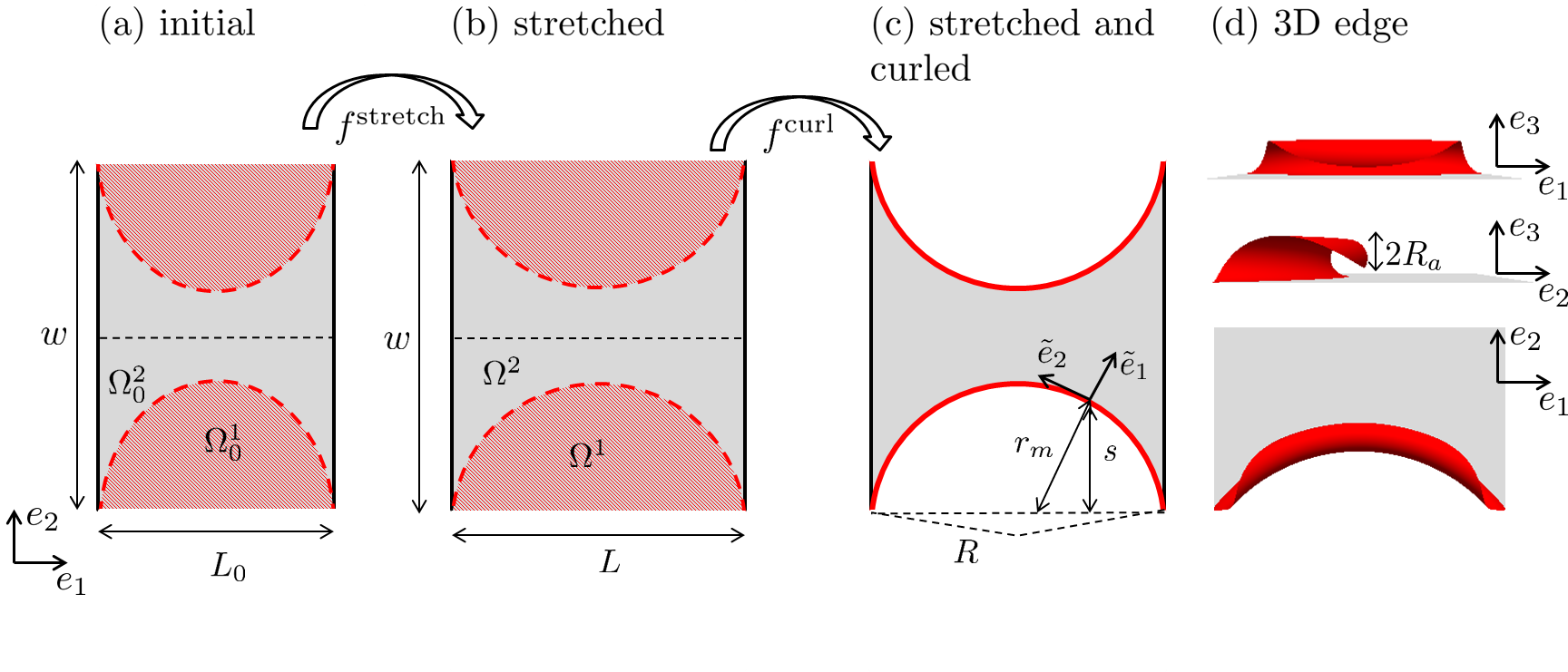}
\caption{\label{fig:curl} Ansatz of the deformation of a flat monolayer from its natural configuration (a) in the absence of active terms (membranal pre-strain and spontaneous curvature) to its curled configuration in presence of activity (c). We decompose the displacement into two steps, $f^{\text{stretch}}$ and $f^{\text{curl}}$. $f^{\text{stretch}}$ brings the monolayer to its actual in-plane dimensions (b) and $f^{\text{curl}}$ curls the margins to reach the current configuration (c). The hatched regions therefore indicate  the material points corresponding to the curled margins in the current configuration. The last panel (d) shows the 3D projections of the ansatz, focusing on the bottom free edge only.}
\end{figure*}

To make this reasoning  quantitative but keep analytical computations tractable, we postulate that the deformation field is an isotropic planar stretch of the rectangular configuration $\Omega_0=(0,L_0) \times (0,w)$ corresponding to a relaxed state in the absence of external stretch and activity into a rectangular configuration with the actual size $(0,L) \times (0,w)$.  This configuration is then combined with a curling normal to the free margins of the monolayer with the constant radius of curvature $R_a$. A more refined ansatz would take into account some expected \cite{callan2012self} self-similar curling at the margin. The shape of the free interface is assumed to be an arc circle of radius $R$. See Fig.~\ref{fig:layer} and \ref{fig:curl}. Given the symmetry of the problem, we only consider the lower half of the monolayer in the following analysis. 

The initial isotropic stretching is related to the deformation ansatz in the $(e_1,e_2,e_3)$ frame:
$$
\left\lbrace\begin{array}{c}
f^{\text{stretch}}_1=x_1\frac{L}{L_0}\\
f^{\text{stretch}}_2=x_2\\
f^{\text{stretch}}_3=0.
\end{array} \right.
$$
Next, the lower edge curling is captured in the Frenet frame $(\tilde{e}_1,\tilde{e}_2,e_3)$ attached to the free margin (See Fig.~\ref{fig:curl}~(c))  by
$$
\left\lbrace\begin{array}{c}
f^{\text{curl}}_1=R_a\sin\left( \frac{\tilde{x}_1}{R_a}\right) \\
f^{\text{curl}}_2=0\\
f^{\text{curl}}_3=R_a\left( 1-\cos\left( \frac{\tilde{x}_1}{R_a}\right) \right).
\end{array} \right.
$$
The curling is normal to the free margin (direction $\tilde{e}_1$) and encompasses the material points denoted as $\Omega^1$ on Fig.~\ref{fig:curl}. Thus $\tilde{x}_1\in [-r_m,0]$, where $r_m$ is the length of material curled at a given point of the interface. \textcolor{black}{A more refined ansatz involving a non constant curvature of the free margin would modify the expression of $f^{\text{curl}}$.} Points outside of the domain $\Omega^1$ are unaffected by the curling. The final deformation is then the composition of the two deformations specified above: $f=f^{\text{curl}}\circ f^{\text{stretch}}$.

Based on $f$, we need to evaluate the total energy $U$ in the reference configuration. To this end, we separately define the deformation into the two domains $\Omega^2$ (where there is no curling) and $\Omega^1$ (where there is curling) in the current configuration and we map them back into $\Omega^2_0$ and $\Omega^1_0$ in the reference configuration. We parametrize $(x_1,x_2)\in\Omega^2_0$ by \textcolor{black}{using a mapping $j_2$ transforming $[0,1]^2$ to $\Omega^2_0$} 
$$j_2:(\lambda_1,\lambda_2)\mapsto \left(x_1=L_0\lambda_1,x_2=\frac{w}{2}+\lambda_2\left[s(\lambda_1)-\frac{w}{2} \right]\right),$$
where variables $\lambda_1$ and $\lambda_2$ vary in the unit interval $(\lambda_1,\lambda_2)\in [0,1]^2$. The expression of the local deflection $s$ (see Fig.~\ref{fig:curl}~(c)) is given by,
$$s(\lambda_1)=\frac{L \sqrt{1-(2 \lambda_1-1)^2 \xi^2} \left(1-\sqrt{\frac{1-\xi^2}{1-(2 \lambda_1-1)^2 \xi^2}}\right)}{2 \xi}.$$
In the above formula, 
\begin{equation}\label{e:xi_def}
 \xi=\frac{L}{2R}
 \end{equation}
is a convenient non-dimensional quantity ranging between $0$ and $1$ that parametrizes the deflection at the center of the layer:
\begin{equation}\label{e:deflection}
d=\frac{L \left(1-\sqrt{1-\xi^2}\right)}{2 \xi}-R_a \sin \left(\frac{L \left(1-\sqrt{1-\xi^2}\right)}{2
   R_a \xi}\right).
\end{equation}
Next, we parametrize the domain $(x_1,x_2)\in\Omega^1_0$ using polar coordinates \textcolor{black}{mapping}, 
\begin{multline*}
j_1:(r,\theta)\mapsto \left(x_1=\frac{L_0}{2}  \left(\sin (\theta ) \left(\frac{2 r}{L}+\frac{1}{\xi}\right)+1\right),\right. \\ 
\left. x_2=\frac{ \cos (\theta ) (L+2 r \xi)-L \sqrt{1-\xi^2}}{2  \xi}\right).
\end{multline*}
The angle $\theta$ thus varies in the range $[-\theta_m,\theta_m]$ where $\theta_m=\arcsin(\xi)\in [0,\pi/2]$ and the radius $r$ varies in the range $[-r_m(\theta),0]$ where $r_m(\theta)=L (1-\sqrt{1-\xi^2} \sec (\theta ))/(2 \xi)$.
When $\xi=0$ (i.e. $\theta_m=0$), the margin is flat and uncurled while the deflection is maximal when $\xi=1$ (i.e. $\theta_m=\pi/2$).

\begin{figure}[h!]
\centering
\includegraphics[width=0.45\textwidth]{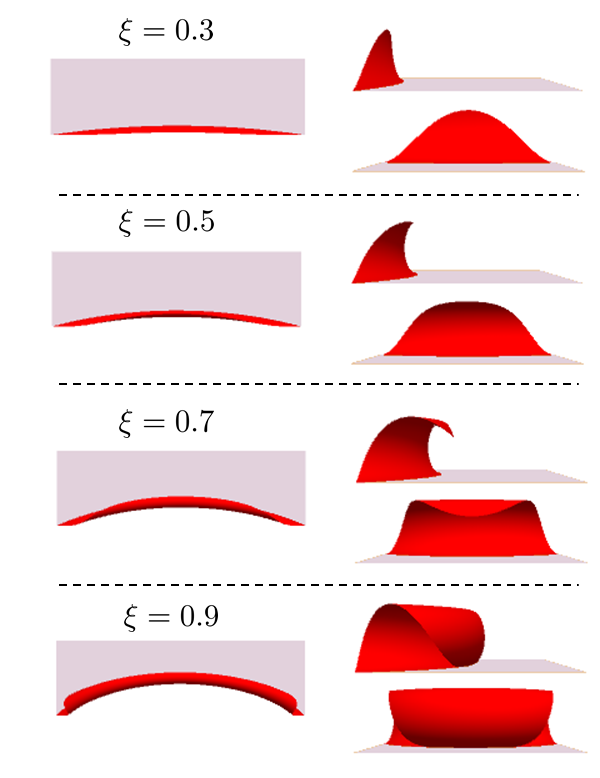}
\caption{\label{fig:xi} Different views of the margin curling when the parameter $\xi$ increases with $L$, $w$ and $R_a$ fixed.}
\end{figure}

In each domain, the final deformation can then be expressed in the  $(e_1,e_2,e_3)$ frame. Namely, deformation in $\Omega^2_0$ is:
\begin{equation}\label{eq:deformation_in_2}
\left\lbrace\begin{array}{c}
f_1=L\lambda_1\\
f_2=\lambda_2 \left(s(\lambda_1)-\frac{w}{2}\right)+\frac{w}{2}\\
f_3=0.
\end{array} \right.
\end{equation}
and in $\Omega^1_0$ it reads,
\begin{equation}\label{eq:deformation_in_1}
\left\lbrace\begin{array}{c}
f_1=\frac{L \sin (\theta )}{2 \xi}+\frac{L}{2}+R_a \sin (\theta ) \sin \left(\frac{r}{R_a}\right)\\
f_2=\cos (\theta ) \left(\frac{L}{2 \xi}-\frac{L \sqrt{1-\xi^2} \sec (\theta )}{2 \xi}\right)+R_a \cos (\theta ) \sin \left(\frac{r}{R_a}\right)\\
f_3=R_a \left(1-\cos \left(\frac{r}{R_a}\right)\right) .
\end{array} \right.
\end{equation}
With $L$, $w$ and $R_a$ given, the deformation $f$ is fully characterized by $\xi$ which controls the amount of curling at the margin as we illustrate on Fig.~\ref{fig:xi}.
With our ansatz of $f$, we can now compute the total elastic energy of the layer 
\begin{multline}
\bar{U}=\frac{U}{2}=\int_{\Omega_0^1}\left[ u_{\text{s}}(E-E_a)+u_{\text{b}}(K-K_a)\right] dx_1dx_2+\\
\int_{\Omega_0^2}\left[ u_{\text{s}}(E-E_a)+u_{\text{b}}(K-K_a)\right] dx_1dx_2 .
\end{multline}
For this we use the deformation fields \eqref{eq:deformation_in_2} and \eqref{eq:deformation_in_1} to compute the strain tensor and the curvature tensor according to formulas \eqref{eq:deformation}-\eqref{eq:Cauchy_strain}-\eqref{eq:Curvature}. The integrals and partial derivatives entering in all formulas have to be computed using the mappings $j_1$ and $j_2$ respectively in $\Omega_0^1$ and $\Omega_0^2$. All these computations, though giving potentially lengthy expressions can be carried out explicitly except for the final integration of local stretching and bending energies.

However, this last step can also be made explicit using the fact that $e$ is very small compared to all other lengths (there are two orders of magnitude between $e\simeq 10\mu$m and $L\simeq 1$mm or $w\simeq 1$ mm)  but spontaneous curvature is very large, essentially of the order of the inverse of a cell thickness . We therefore assume that $R_a$ is of the form 
$$R_a=\frac{e}{k_a},$$
where $k_a$ is a non-dimensional parameter quantifying the magnitude of the spontaneous curvature \cite{hannezo2014theory}. The implication of this last assumption is that, despite the smallness of the thickness $e$, the stretching and bending energies have terms contributing at first order in $e$ and can locally balance each other even in the limit where $e$ is vanishingly small. Indeed, both \eqref{eq:deformation_in_2} and \eqref{eq:deformation_in_1} can generically be written as
$$f(x)=f^0(x)+e f^1\left( x,\frac{x}{e}\right) $$ 
such that the bending term in \eqref{e:elastic_energy} contributes to $U$ through $f^1$ even at first order in $e$. The computation of $\bar{U}$ at first order in $e$ essentially involves asymptotic expansions in the small parameters $e/L$ and $e/w$ as well as some averaging over the variable $x/e$ which varies very quickly compared to $x$ (similar to the technique employed in vibrational mechanics \cite{blekhman2000vibrational} to extract the slow part of the motion).

Note that the average tension on the cantilevers only involves the elementary deformation field in $\Omega^2$ and thus takes a simple form  
$$T=\frac{e Y}{6} \lambda_L^2  \left(\lambda_L^2-2 \epsilon_a+1\right) \left(1+\lambda_L^2-4 \epsilon_a-2\right),$$
where the stretch variable 
$$\lambda_L=L/L_0,$$
can be experimentally adjusted  by moving the cantilevers. For small applied strains the tension reads,
$$T=T_a+e Y_a(\lambda_L-1)$$
where the active stress reads, $T_a= 4 e Y (\epsilon_a-1)\epsilon_a/3$  and the active rigidity is $Y_a=2 Y \left(4 \epsilon_a^2-7 \epsilon_a+1\right)/3$. It is known that the effective stiffness of a single cell \cite{etienne2015cells} or a tissue \cite{vincent2015active,wyatt2019actomyosin} can change  when some of its molecular motors are inhibited or activated.

The value of $\epsilon_a$ was measured in \cite{wyatt2019actomyosin} to be $\epsilon_a\simeq -0.4$ and the effective modulus $Y_a \simeq 650$ Pa.

\section{Initial deflection}\label{sec:initial_deflection}

Before any stretching is applied, we consider that $\lambda_L=1$ such that the only source of tension $T=T_a>0$ in the monolayer is endogenous. This assumption was experimentally checked with drugs inhibiting the motors activity in \cite{wyatt2019actomyosin}.

\begin{figure}[h!]
\centering
\includegraphics[width=0.45\textwidth]{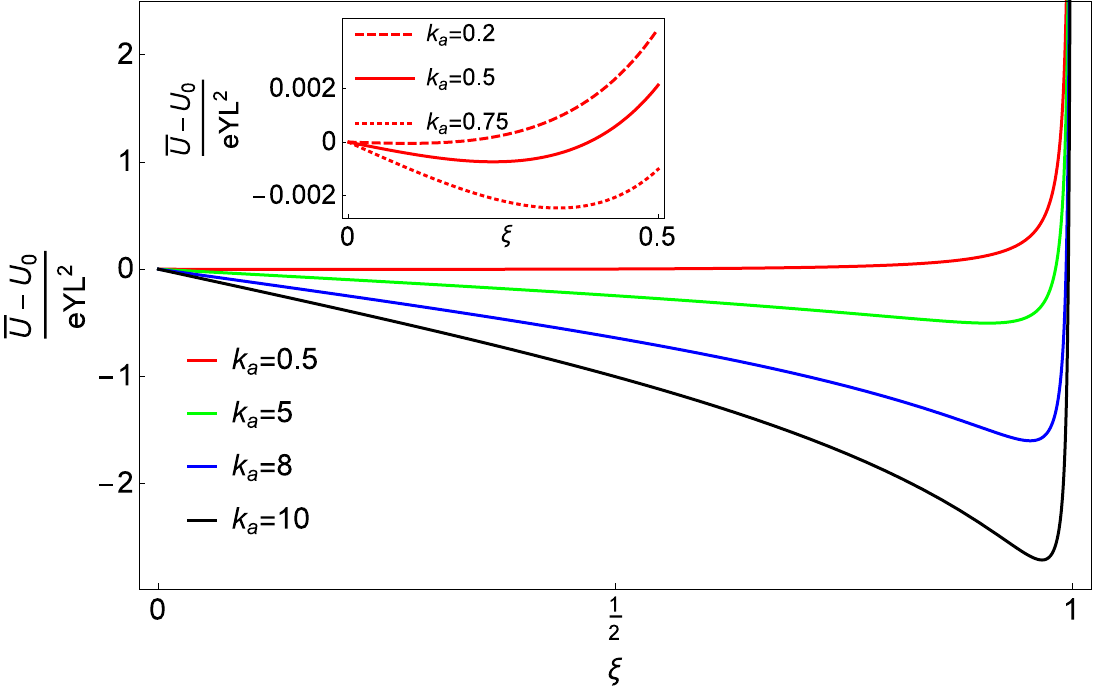}
\caption{\label{fig:energyx} Dependence of the energy $U$ on the deflection variable $\xi$ for a given value of $\epsilon_a=-0.4$. The inset shows the decay of the energy with $\xi$ for experimentally relevant values of $k_a\sim 1$ \cite{Fouchard806455}}
\end{figure}

At the leading order in the monolayer thickness, the energy then takes the form,
\textcolor{black}{\begin{multline*}
\frac{\bar{U}-U_0}{eYL^2}=\frac{1}{144 \xi^2
   \left(1-\xi^2\right)}\left[\left(-4 k_a^2 \left(1-\xi^2\right)+36 \left(1-\xi^2\right) \epsilon_a+3\right)\times\right.\\
   \left.\left(\sin ^{-1}(\xi)-\xi \sqrt{1-\xi^2}\right)
   +12 \left(\xi^2-1\right) (3 \epsilon_a+1) M(\xi)+6  \xi^3\sqrt{1-\xi^2}\right]  
\end{multline*}}
where the special function $M$ can be expanded in power series,
\begin{multline}
M(\xi)=\sum_{k=1}^{\infty}(-1)^k \left(\frac{2 \sin^{-1}(\xi) \cos (2 k \sin^{-1}(\xi))}{k}\right.\\
-\left.\frac{\sin (2 k \sin^{-1}(\xi))}{k^2}\right).
\end{multline}
and $U_0=\bar{U}(\xi=0)$ is a constant independent of $\xi$. We show on Fig.~\ref{fig:energyx} the typical behavior of the energy for several values of $k_a$. 
For small values of $\xi$, it decreases as 
$$\bar{U}-U_0\underset{\xi\rightarrow 0}{\sim} - \frac{eYL^2 k_a^2}{54}\xi$$
because curling more material reduces the bending energy while the stretching energy is negligible. This is a bending dominated regime. However, in the opposite limit where  $\xi$ approaches $1$ the energy diverges when as 
$$\bar{U}-U_0\underset{\xi\rightarrow 1}{\sim} \frac{eYL^2\pi}{192(1-\xi)}$$
and is dominated by the stretching contribution. The balance between these two regimes determines the equilibrium shape of the free margin of the monolayer.

\begin{figure}[h!]
\centering
\includegraphics[width=0.3\textwidth]{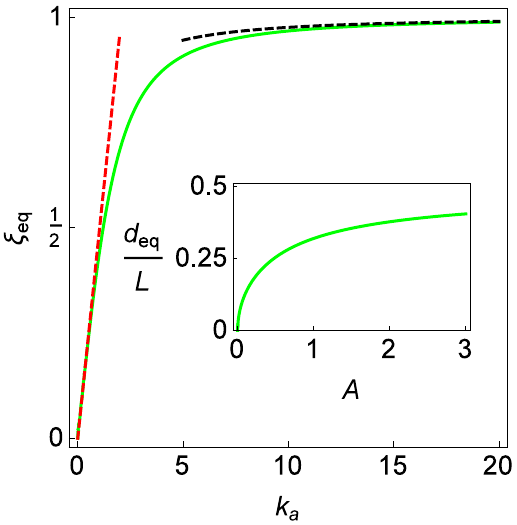}
\caption{\label{fig:xeqinit} The full green line shows the equilibrium value of $\xi$ as a function of $k_a$ obtained by numerical minimization of the potential energy $\bar{U}$. The dashed lines correspond to the asymptotic formulas \eqref{e:x_small} (in the $k_a\ll 1$ regime) and \eqref{e:x_large} (in the $k_a\gg 1$ regime). Parameter $\epsilon_a=-0.4$. The inset shows how the deflection associated to the value of $\xi_{\text{eq}}$ varies as a function of the motor activity which proportionally affects both the pre-strain and the spontaneous curvature (see \eqref{eq:activity}). Parameters $\bar{\epsilon}_a=-0.4$ and $\bar{k}_a=5$.}
\end{figure}

To find the balancing point between bending and stretching, we therefore minimize $\bar{U}$ with respect to $\xi$.  It is important to note that $Y$ drops out from the minimization and therefore does not influence $\xi_{\text{eq}}$, the equilibrium value that minimizes the energy. We show on Fig.~\ref{fig:xeqinit} the dependence of $\xi_{\text{eq}}$ on  the spontaneous curvature parameter $k_a$.
As we expect, $\xi_{\text{eq}}$ increases with $k_a$ as an increase of spontaneous curvature favors curling. The dependence of $\xi_{\text{eq}}$ on $k_a$ can be analytically captured in two asymptotic cases. When $k_a\ll 1$ we obtain,
\begin{equation}\label{e:x_small}
\xi_{\text{eq}}\sim \sqrt{\frac{5}{6}}\frac{k_a}{3\sqrt{-\epsilon_a}},
\end{equation}
which degenerates as a square root dependence when $\epsilon_a=0$: $\xi_{\text{eq}}\sim (7/3)^{1/4}\sqrt{k_a/2}$. Interestingly, this limit still accurately captures the value of $\xi_{\text{eq}}$ up to moderate values of $k_a\lesssim 1$ that correspond to some experimental measurements \cite{Fouchard806455}.  In the opposite limit when $k_a\ll 1$, we can approximate $\xi_{\text{eq}}$ by:
\begin{equation}\label{e:x_large}
\frac{24 \pi }{(1-\xi_{\text{eq}})^2}\sim \frac{128 \sqrt{2} \left(k_a^2-18 \epsilon_a-4\right)}{\sqrt{1-\xi_{\text{eq}}}}+\frac{288 \pi  (2 \epsilon_a+1)}{1-\xi_{\text{eq}}}.
\end{equation}
Note that when $k_a>1$, the spontaneous radius of curvature is larger than the monolayer thickness which is admissible since it is induced by a mismatch of apical and basal tension \cite{hannezo2014theory} and not related to the actual curvature of a single cell. However, the applicability of our ansatz may be questioned in this case where mechanical contacts between the folds of the curled region may play an important role. The account of such non-penetration constraints would require a complex numerical treatment. 

The equilibrium value of $\xi_{\text{eq}}$ can be easily translated into a measured deflection through equation \eqref{e:deflection} which, at zeroth order in $e/L$ reads,
$$\frac{d}{L}=\frac{ 1-\sqrt{1-\xi^2}}{2 \xi}$$ 
and increases from $0$ when $\xi=0$ to $1/2$ when $\xi=1$. 
\begin{figure}[h!]
\centering
\includegraphics[width=0.4\textwidth]{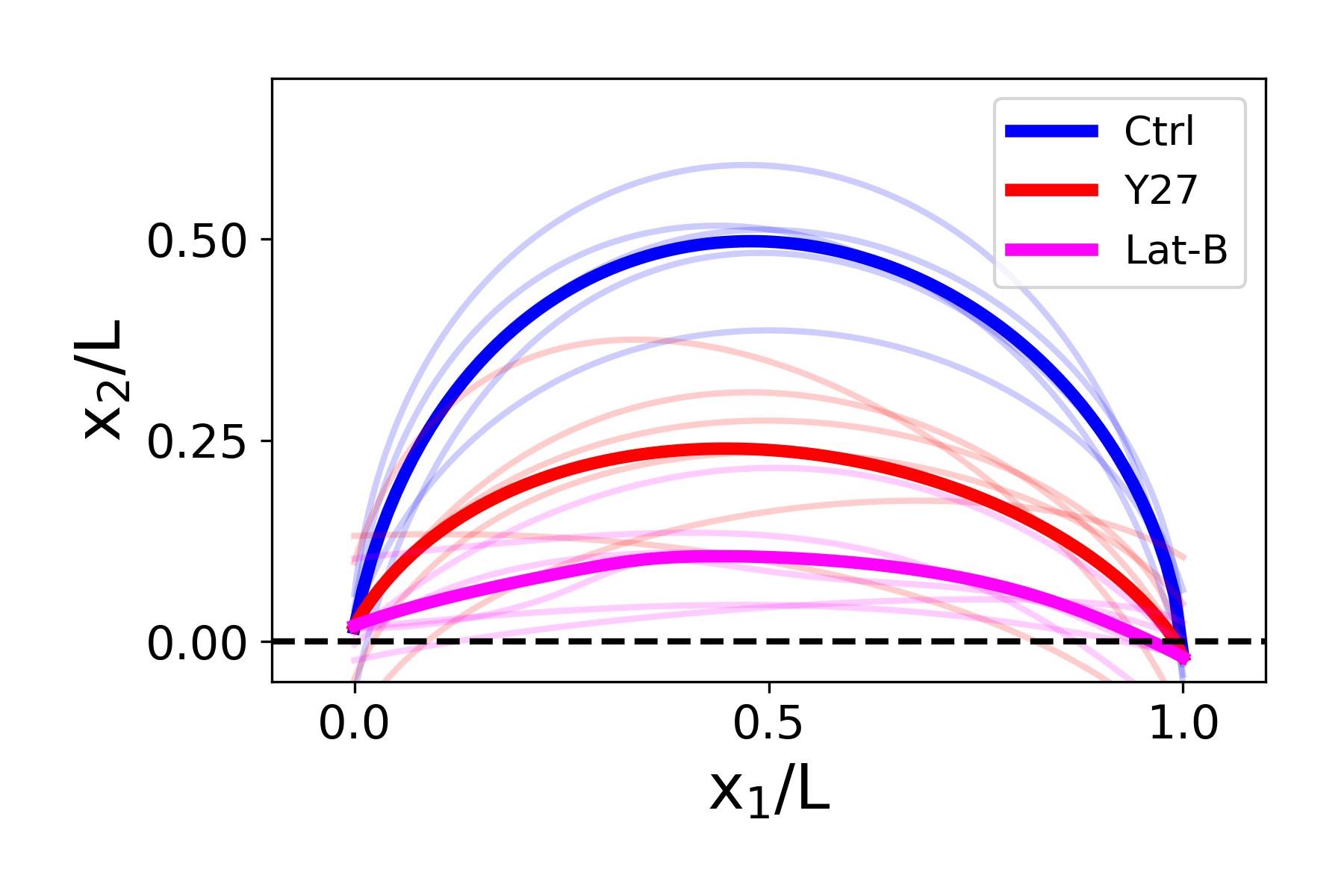}
\caption{\label{fig:pharma_treatments}  Shape of the free edge of monolayers ($\Gamma_l$ on Fig.\ref{fig:layer}) in a control untreated situation (blue line, N=5) and with two drug treatments (Y27, red line, N=6) and (Latrunculin-B, purple line, N=6) that impair the active tensions in the cells cortex. The thick lines represent the mean behavior while the light lines are directly extracted from experiments. The dashed black line corresponds to a straight bridge between the two cantilevers. \textcolor{black}{Misalignment of the monolayers boundaries appears as a result of the uneven spreading of the monolayers on each of the two plates.}}
\end{figure}
It is  not directly obvious how an inhibition or promotion of molecular motors activity will affect the deflection because motors control both $\epsilon_a$ and $k_a$ which can have antagonistic roles on $\xi_{\text{eq}}$ (see \eqref{e:x_small}).  However, a simple assumption is that $\epsilon_a$ scales with the average of the activity of the motor on the apical and basal side of the monolayer while $k_a$ scales with the difference between the activities on both faces of the monolayer. In this respect, it is reasonable to assume that both $k_a$ and $\epsilon_a$ are affected in the same proportion if the motor activity is modified genetically or with drugs. We formally express this proportionality as
\begin{equation}\label{eq:activity}
k_a=\bar{k}_aA\text{ and } \epsilon_a=\bar{\epsilon}_aA,    
\end{equation}
where $A$ is a non-dimensional measure of the motor activity and we show on the inset of Fig.~\ref{fig:xeqinit} the dependence of $d_{\text{eq}}/L$ on $A$. We observe that the effect of spontaneous curvature surpasses that of in-plane pre-stress to increase the deflection of the margin when activity increases. In agreement with this trend, we show on Fig.~\ref{fig:pharma_treatments} the equilibrium shape of the monolayer free edge in response to two pharmacological treatments that reduce the activity of the cell monolayer either by partially inhibiting the molecular motors (Y27 curve) or by partially depolymerizing the polymers that serve as scaffolds for molecular motor contractility (Lat B curve).

\section{Deflection to elongation relation}\label{sec:deflection_elongation}

\begin{figure}[h!]
\centering
\includegraphics[width=0.4\textwidth]{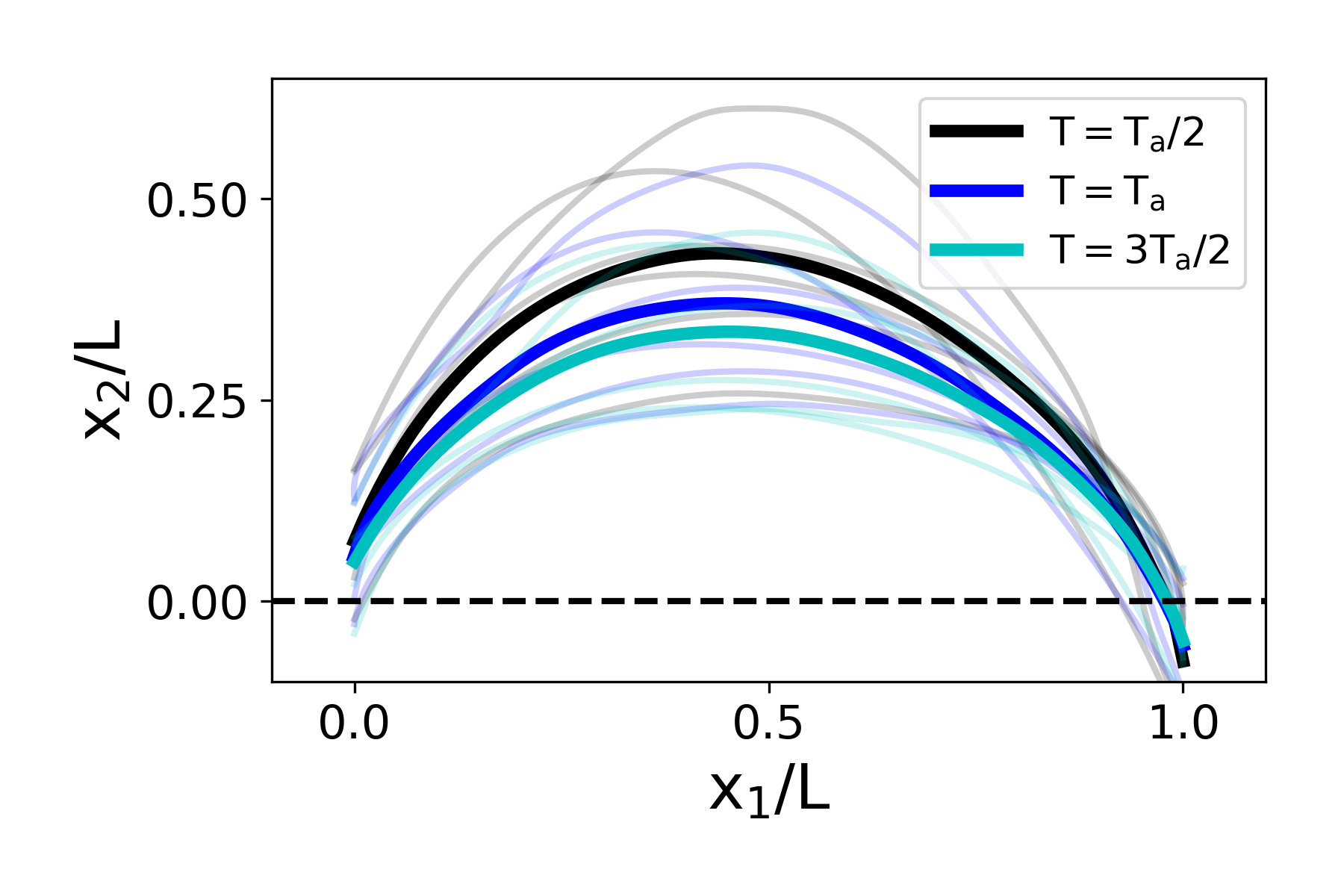}
\caption{\label{fig:tension_shape}  Shape of the free edge of monolayers ($\Gamma_l$ on Fig.\ref{fig:layer}) for different applied tensions. The blue line corresponds to the initial state where $\lambda_L=0$ and $T=T_a$. The black line corresponds to a smaller tension $T=\frac{T_a}{2}$. The cyan line corresponds to a larger tension  $T=\frac{3T_a}{2}$.  The thick lines represent the mean behavior while the light lines are directly extracted from experiments (N=6). The dashed black line corresponds to a straight bridge between the two cantilevers. \textcolor{black}{Misalignment of the monolayers boundaries appears as a result of the uneven spreading of the monolayers on each of the two plates.}}
\end{figure}

From the initial configuration, we can experimentally apply a finite stretch to the mobile cantilever and observe that the deflection decreases, see Fig.~\ref{fig:tension_shape}, while we would expect an increase of the necking for a passive elastic sheet. To rationalize this observation, we can compute again the elastic energy which takes a more complex form in this case (see expression in Appendix.~\ref{sec:appendix_A}) 

\begin{figure}[h!]
\centering
\includegraphics[width=0.45\textwidth]{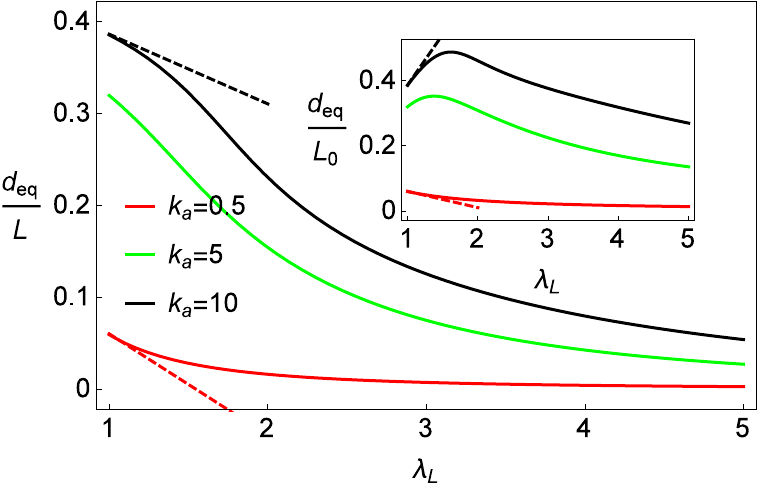}
\caption{\label{fig:deflection_stress} Deflection of the layer margin as a function of the stretch. The dashed lines are related to the asymptotic formulas \eqref{e:deflection_stress_small} (when $k_a\ll 1$, red) and \eqref{e:deflection_stress_large} (when $k_a\gg 1$, black). We show in inset the value of the deflection normalized by the initial length $L_0$ instead of the actual length. Parameter $\epsilon_a=-0.4$}
\end{figure}

As for the initial case, this form has a single minimum in $\xi$ corresponding to the equilibrium deflection of the free tissue margin. Fig.~\ref{fig:deflection_stress} shows how the deflection depends on the applied stretch for small, moderate and large values of $k_a$.

For a small value of $k_a\ll 1$, we can compute the deflection for small strains
\begin{equation}\label{e:deflection_stress_small}
\frac{d_{\text{eq}}}{L}\sim \frac{\sqrt{\frac{5}{6}} k_a}{12 \sqrt{-\epsilon_a}}-\frac{\sqrt{\frac{5}{6}} (\lambda_L -1) k_a}{36 (-\epsilon_a)^{3/2}}
\end{equation}
and in the $k_a\gg 1$ regime we obtain,
\begin{equation}\label{e:deflection_stress_large}
\frac{d_{\text{eq}}}{L}\sim\frac{1}{2}-\frac{\sqrt[3]{3 \pi }}{4k_a^{2/3}}  -\frac{\sqrt[3]{\pi } (\lambda_L -1) }{2\ (3k_a)^{2/3}}.
\end{equation}
While the value of the deflection itself is larger for a higher motor activity $A$, we expect the slope of the deflection   under stretch  to decrease with the motor activity, because such slope scales as $A^{-1/2}$ for $A\ll 1$ (see \eqref{e:deflection_stress_small}) and $A^{-2/3}$ for $A\gg 1$ (see \eqref{e:deflection_stress_large}). 

We also show in inset of Fig.~\ref{fig:deflection_stress} the value of the deflection normalized by the initial length instead of the actual length. Note that unlike $d_{\text{eq}}/L$, $d_{\text{eq}}/L_0$ does not have to be smaller than $1/2$. In the large spontaneous curvature regime, this ratio starts to increase for  small strains. This is because $\xi_{\text{eq}}$ remains close to $1$ since the energy needed to uncurl the margin is very large while the length increases. Ultimately, as the stretch becomes large, it can again balance the bending energy and uncurls the margin leading to a decrease of $d_{\text{eq}}/L_0$. This behavior is not present in the $k_a\ll 1$ regime where $d_{\text{eq}}/L_0$ immediately starts to decrease by uncurling the margin in response to even small stretches. \textcolor{black}{Note that the large tension limit is not necessarily well captured by our ansatz since it does not account for stress concentration phenomena involved in necking and/or mechanical damage of the cell monolayer under loading. }

\section{Discussion}\label{sec:discussion}

We begun by studying the case of the buckling of an elastic film suspended between two cantilevers and subjected to in-plane strain to illustrate the fact that there is a transition from a regime dominated by the stretching energy when the film is put under tension to a regime dominated by the bending energy when the film is compressed beyond a critical threshold. While the two energies do compete to set the value of this buckling threshold, only one of the two is important in each regime to determine at least qualitatively the object shape. Another signature in this passive case is that buckling does not happen continuously as the compression is gradually increased but suddenly through a bifurcation at the critical loading threshold. 

Next, to model the activity of the cellular monolayer, following the framework of NEP, we have augmented the passive film model by introducing a spontaneous in-plane contractility and out-of-plane curvature that originate from the presence of molecular motors unevenly distributed along the film thickness. As a result, the spontaneous curvature scales with the inverse of the film thickness leading, even in the absence of an external loading, to a competition between the stretching and bending energies to set the shape of the free edge of the film. 

More precisely, by assuming that the shape of the free margin is an arc of a circle, the elastic energy depends on only a single free parameter that quantifies the central deflection of the film. We then show that the minimum of the energy corresponding to the mechanical equilibrium of the film exhibits a deflection that balances stretching and bending. We obtain the expression of this deflection as a function of the active parameters quantifying the contractility and spontaneous curvature and conclude, in agreement with experiments, that increasing the molecular motor activity leads to a larger deflection.

Interestingly, increasing the external stretch applied to the monolayer continuously modifies the balancing point between stretching and bending in a non-trivial manner. If the spontaneous curvature is not too high, the prevailing effect is to uncurl the tissue margin leading to a decrease of the deflection as observed in experiments. However, in the limit of a high spontaneous curvature, we predict that the deflection will first increase as for a passive material because uncurling the layer requires a lot of energy until the stretching is enough to uncurl the margin and the deflection decreases again.
 
Overall, our results suggest that unlike in the case of passive slender elastic objects where the transition from bending to stretching happens through a sharp transition when the loading is changed, the presence of a spontaneous curvature scaling with the inverse of the film thickness  leads to a competition between stretching and bending that is continuously affected by an external loading. Such competition may be crucial to understand some three-dimensional mechanical events that happen during morphogenesis such as the formation of folds and invaginations for instance during gastrulation; or the fracture of an epithelium which happens during the \emph{Drosophila} leg disc eversion \cite{Fouchard806455}.

One interesting follow-up of this work would be to solve the full mechanical problem with the new assumption of a small spontaneous curvature scaling like the inverse of the film thickness formulated above instead of using an ansatz for the deformation. By doing so, one would be able to find the real equilibrium shape of the tissue margin (i.e. not approximating it by an arc of circle) which will be characteristic of the competition between stretching and bending. \textcolor{black}{ Other non-linear effects could also be investigated in this way such as necking under tension or wrinkling \cite{nelson2016buckling} .}  A more fundamental perspective that is suggested by our results is to rigorously develop a theory for elastic plates with a spontaneous curvature that scales with the inverse of the thickness instead of postulating the NEP type energy \eqref{e:elastic_energy} used in this work. This may generalize the framework developed \cite{lewicka2010foppl} that assumes a finite spontaneous curvature.

\textcolor{black}{Another important generalization of this work would be to account for cell-cell rearrangements that are known to happen over a long timescale during many developmental processes \cite{wyatt2016question}, such as convergence and extension \cite{munro2002polarized}. This would require to specify in a self-consistent way the time evolution of the target metric controlled by $E_a$ and $K_a$ as is done for instance in the framework of morpho-elasticity \cite{goriely2017mathematics}. However, in the experiments presented here, cell-cell rearrangements have been shown to be negligible over hour long time-scales \cite{wyatt2015emergence}.} 

\acknowledgments{ P.R. acknowledges support from a CNRS-Momentum grant. J.F. and P.R. were funded by BBSRC grant (BB/M003280 and BB/M002578) to G.C. and A.K. J.F, T.W., N.K. and G.C.were supported by a consolidator grant from the European Research Council to G.C. ( MolCellTissMech, agreement 647186). T.W. and N.K. were funded by the UCL Graduate School and the EPSRC funded doctoral training programme CoMPLEX. N.K. was also in receipt of a UCL Overseas Research Scholarship.}

\appendix

\section{Expression of the elastic energy when $\lambda_L\neq 0$}\label{sec:appendix_A}

\onecolumngrid
\begin{multline}
\frac{\bar{U}-U_0}{eYL_0^2}=\frac{\delta _L}{{1152 \xi^2 \left(\xi^2-1\right)}} \left(\xi \sqrt{1-\xi^2} \left(k_a^2 \left(\left(4 \xi^4-34 \xi^2+30\right) \delta _L^4+\left(-8 \xi^4+92 \xi^2-84\right) \delta _L^2+4 \xi^4-26 \xi^2+22\right)+\right. \right. \\
\left. \left. \delta _L^2 \left(288 \left(\xi^2-1\right) \epsilon _a-8 \xi^4+248
   \xi^2-246\right)-576 \xi^2 \epsilon _a+576 \epsilon _a+\left(4 \xi^4-64 \xi^2+45\right)  \delta _L^4+4 \xi^4-232 \xi^2+225\right)-\sin ^{-1}(\xi)\right.\\
\left.    \left(2 k_a^2 \left(3 \left(4 \xi^4-9 \xi^2+5\right) \delta _L^4+\left(-32 \xi^4+74
   \xi^2-42\right) \delta _L^2+20 \xi^4-31 \xi^2+11\right)+3 \left(-2 \delta _L^2 \left(48 \left(\xi^2-1\right)^2 \epsilon _a+32 \xi^4- 74 \xi^2+\right. \right. \right. \right.  \\   
\left. \left. \left. \left. 41\right)+96 \left(\xi^4-3 \xi^2+2\right) \epsilon _a+\left(24 \xi^4-34 \xi^2+15\right) \delta
   _L^4+40 \xi^4-114 \xi^2+75\right)\right)+12 \left(\xi^2-1\right) \left(\delta _L^2+1\right)  \left(-12 \epsilon _a+\delta _L^2-5\right)M(\xi)\right).
\end{multline}
\twocolumngrid

\end{document}